\documentclass[a4paper,fleqn,usenatbib]{mnras}
\usepackage{newtxtext,newtxmath}
\usepackage[T1]{fontenc}
\usepackage{ae,aecompl}
\usepackage{natbib}

\usepackage{graphicx}
\usepackage{amsmath}
\usepackage{amssymb}

\usepackage{aas_macros}
\usepackage{color}
\usepackage{epsfig}
\usepackage{float}
\usepackage{latexsym}
\usepackage{morefloats}
\usepackage{times}

\title[Environmental Galactic Conformity]{The Large-scale Effect of Environment on Galactic Conformity}

\author[Sun et al.]
{\parbox{\textwidth}
{Shuangpeng Sun,$^{1,2}$\thanks{E-mail: sunshp@nao.cas.cn} 
Qi Guo,$^{1,2}$\thanks{E-mail: guoqi@nao.cas.cn} 
Lan Wang,$^{1}$ 
Jie Wang,$^{1,2}$ 
Liang Gao,$^{1,2,3}$ 
Cedric G. Lacey$^{3}$ 
and Jun Pan$^{1}$ 
\vspace{.20cm}} \\
$^{1}$Key Laboratory for Computational Astrophysics, National Astronomical Observatories, Chinese Academy of Sciences, Beijing, 100012, China\\
$^{2}$School of Astronomy and Space Science, University of Chinese Academy of Sciences, Beijing 10039, China \\
$^{3}$Institute of Computational Cosmology, Department of Physics,
University of Durham, South Road, Durham DH1 3LE, UK}

\date{Accepted XXX. Received YYY; in original form ZZZ}
\pubyear{****}

\begin{document}
\label{firstpage}
\pagerange{\pageref{firstpage}--\pageref{lastpage}}
\maketitle

\begin{abstract}

We use a volume-limited galaxy sample from the SDSS Data Release 7 to explore the dependence of galactic conformity on the large-scale environment, measured on $\sim$ 4 Mpc scales. We find that the star formation activity of neighbour galaxies depends more strongly on the environment than on the activity of their primary galaxies. In under-dense regions most neighbour galaxies tend to be active, while in over-dense regions neighbour galaxies are mostly passive, regardless of the activity of their primary galaxies. At a given stellar mass, passive primary galaxies reside in higher density regions than active primary galaxies, leading to the apparently strong conformity signal. The dependence of the activity of neighbour galaxies on environment can be explained by the corresponding dependence of the fraction of satellite galaxies. Similar results are found for galaxies in a semi-analytical model, suggesting that no new physics is required to explain the observed large-scale conformity.

\end{abstract}


\section{Introduction} 

Star formation is one of the most important mechanisms by which galaxies grow (e.g. \citealt{Guo2008}, \citealt{Vogelsberger2014a}). It can be affected by gas fuelling, feedback, stripping, depletion, etc. Star formation can be quenched in central galaxies mostly by internal mechanisms, such as supernovae feedback and AGN feedback, while for satellite galaxies, their evolution can be significantly affected by environment (e.g. \citealt{Weinmann2009}, \citealt{Peng2012}). 

\cite{Weinmann2006} (hereafter W06) found that the evolution of central galaxies and their satellites related to each other, the so-called "galactic conformity" effect.  For given stellar masses, passive centrals tend to be surrounded by passive satellites, while active centrals to be surrounded by active satellites. In the standard scenario, galaxies form in potential wells dominated by hierarchically merging structures of dark matter. Galaxy properties are closely related to their host dark matter haloes. \cite{Wang2012} found that such galactic conformity results from the fact that red centrals tend to reside in more massive haloes within which the satellite quenching by ram-pressure or tidal stripping is more effective (see also e.g.  \citealt{Peng2010}, \citealt{Henriques2017}). This has been further investigated by \cite{Knobel2015} by removing the halo mass dependence. They found that the conformity signal persists for galaxies within halos of the same masses. It could be that other halo properties, such as halo formation time/assembly bias, also contribute  to the galactic conformity (e.g. \citealt{Gao2005}, \citealt{Wang2013a}).

Lately \cite{Kauffmann2013} (hereafter K13) extended such studies to larger scales, to look at the relation between central galaxies and their galaxy neighbours. Surprisingly, they found that for galaxies as massive as our Milky Way, there exists a strong conformity signal up to 4 Mpc, about 10 times the typical virial radius of the host haloes, suggesting a co-evolution between well separated distinct haloes. One possible explanation is that those co-evolving distinct systems share a common large-scale structure, which has a significant impact on their halo/galaxy properties. For example, haloes in high-density large-scale environments tend to form earlier and thus host more passive galaxies (e.g. \citealt{Thomas2005}, \citealt{Nelan2005}, \citealt{Wang2013a}). However, recent semi-analytic models have taken this effect into account, and K13 found this effect is not enough to explain such a strong conformity signal. \cite{Kauffmann2015} found an excess of massive galaxies with $M_*>10^{11.3}M_{\odot}$ around passive primaries and they tend to have higher possibility to host radio-loud active galactic nuclei (AGN). Therefore it was suggested that the suppression of star formation rates (SFRs) in neighbour galaxies around passive primaries is likely caused by AGN feedback on large scales. However, such effect was not found in the Illustris cosmological hydrodynamical simulation \citep{Vogelsberger2014}, which includes AGN feedback in their models. \citet{Tinker2017} showed that the strong large-scale conformity signal in K13 is almost entirely eliminated by removing a small number of satellites which are misclassified as centrals. \cite{Sin2017} claimed that the strong conformity signal is mainly produced by particular methodologies, so that the signal originates mainly from a very small number of central galaxies in the vicinity of a few very massive clusters.  \citet{Zu2018} confirmed the \cite{Tinker2017} and \citet{Sin2017} results that the strong two-halo conformity signal is primarily due to a misclassification of central galaxies.

This misclassification of satellite galaxies as central galaxies and the environmental effect could be highly degenerate. For example, the possibility to misclassify satellite galaxies as central galaxies could be significantly higher in high-density regions. Nevertheless, in high-density regions, a significant fraction of real central galaxies could also be missed. The relation between the effects caused by the misclassification and the effects caused by environment are thus non-trivial.  It is difficult to distinguish centrals from satellite galaxies with high purity and completeness simultaneously.  Here we take another perspective, focusing on how the large-scale environments affect the conformity signal rather than exploring how good a particular set of selection criterion is in isolating central galaxies.  The paper is organized as follows. Section 2 details the observational spectroscopic sample from the NYU-VAGC catalogue and the semi-analytic model data, and the definition of environment. Section 3 presents how the large-scale conformity signal behaves within different environments as well as the comparison with results from galaxy formation models. Conclusions are summarized in section 4.

\section{Data and methodology} 

\subsection{Observational data} 
\label{sec:observ_data}

We select a volume-limited sample of 25944 galaxies from the spectroscopic New York University Value-Added Galaxy Catalogue (NYU-VAGC; \cite{Blanton2005}) constructed from Data Release 7 of the Sloan Digital Sky Survey \citep{Abazajian2009}. It covers a redshift range of 0.017 < z < 0.04 and a mass range of log ${\rm M}_*$/${\rm M}_{\odot}$ > 9.6. The stellar masses are taken from the NYU-VAGC k-correction catalogue, derived according to \cite{Blanton2007}, and the star formation rates are derived according to \cite{Brinchmann2004}\footnote{http://wwwmpa.mpa-garching.mpg.de/SDSS/DR7/}. 

This sample is similar to but slightly different from the sample used by K13. It is limited to a stellar mass higher by 0.35 dex and covers a larger volume by a factor of 2.7; the star formation rate refers to the total star formation rate both for primary galaxies and neighbour galaxies, rather than the values estimated within the SDSS fibre aperture. K13 used the total SFR, fibre-aperture SFR, HI gas fraction and HI-deficiency to split their primary galaxies into different activity levels and found there was not much dependence of the conformity signal on different indicators. We therefore use only the total SFR to split our primary samples.

Similarly to \citet{Kauffmann2013}, \citet{Tinker2017} ,\citet{Sin2017} and \citet{Zu2018}, we define a galaxy as a primary galaxy if there are no bright galaxies in its vicinity. As discussed in \cite{Sin2017,Tinker2017,Zu2018} the contamination from satellite galaxies could be a serious issue for the study of conformity. Here we adopt a similar but more restricted isolation criterion than K13, i.e. a galaxy with stellar mass ${\rm M}_*$ is identified as a primary galaxy if there is no other galaxy with stellar mass greater than ${\rm M}_*$/2 within a projected radius of ${\rm R_{proj}} $ = 500 kpc and with velocity difference less than 1000 km/s. This velocity difference is larger by a factor of two than that adopted by K13. In total, we have 13415 primary galaxies. The remaining galaxies that do not satisfy the criterion are referred to as satellite galaxies. Neighbour galaxies are defined as those within ${\rm R_{proj}} <$ 4 Mpc and with velocity difference less than 1000 km/s with respect to their primaries. When a primary galaxy is found in the vicinity of another primary galaxy, it is defined as the primary neighbour galaxy.

 In this paper we apply a flat $\Lambda$CDM cosmology model with $\Omega_m$ = 0.3, $\Omega_{\Lambda}$ = 0.7, and $H_0$ = 70 km s$^{-1}$Mpc$^{-1}$ for the analysis of the observational data.

\subsection{Semi-analytic galaxy catalogue} 
\label{sec:SAM_data}

In order to confront the model prediction with observations, we apply an identical analysis to the simulated galaxy catalogue of \cite{Guo2011} (hereafter G11). This catalogue was generated by implementing semi-analytical galaxy formation models (SAMs) on merger trees extracted from a N-body cosmological simulation, the Millennium-II Simulation \citep[MSII,][]{Boylan-Kolchin2009}. The MS-II adopts a flat $\Lambda$CDM cosmology model with  parameters of $\Omega_m$ = 0.25, $\Omega_b$=0.045,  $\Omega_{\Lambda}$ = 0.75, \textit{n}=1, $\sigma_8$=0.9, and $H_0$ = 73 km s$^{-1}$Mpc$^{-1}$. It traced $2160^3$ particles in a box of 100  Mpc/h on each side. The mass of each dark matter particle is 6.88 $\times$ $10^6$ $\rm M_\odot$/h. The particle data were stored at 68 logarithmically spaced snapshot outputs from z = 127 to z =0. At each snapshot, particles are linked together to form a friends-of-friends (FOF) group if their separation is smaller than 0.2 times the mean inter-particle separation ~\citep{Davis1985}. Galaxies at the bottom of the potential well of the FOF group are defined as central galaxies, while other galaxies within the FOF groups are satellite galaxies. Hot gas that got shock-heated during the infall of the IGM can cool down on to the central galaxies and fuel star formation. The central galaxies can also grow by accretion of their surrounding galaxies within the FOF group. When central galaxies fall into even larger systems, they become satellite galaxies. Much less gas can fall onto satellite galaxies due to the environmental effects, and star formation ceases within a few gigayears. More details of the models and the simulations can be found in G11 and \citealt{Boylan-Kolchin2009}. 

As discussed above, the observational selection criterion could misclassify some satellite galaxies as central galaxies, and vice versa. 
To avoid such notational confusion, hereafter we use "primary" and "satellite" galaxies to refer to galaxies defined using the selection criterion, and use "central-SAM"/"satellite-SAM" to refer to the "true" central/satellite galaxies identified in the simulation.

\subsection{Definition of environments}
\label{sec:abund}

\begin{figure}
\centering
\includegraphics[width=\linewidth]{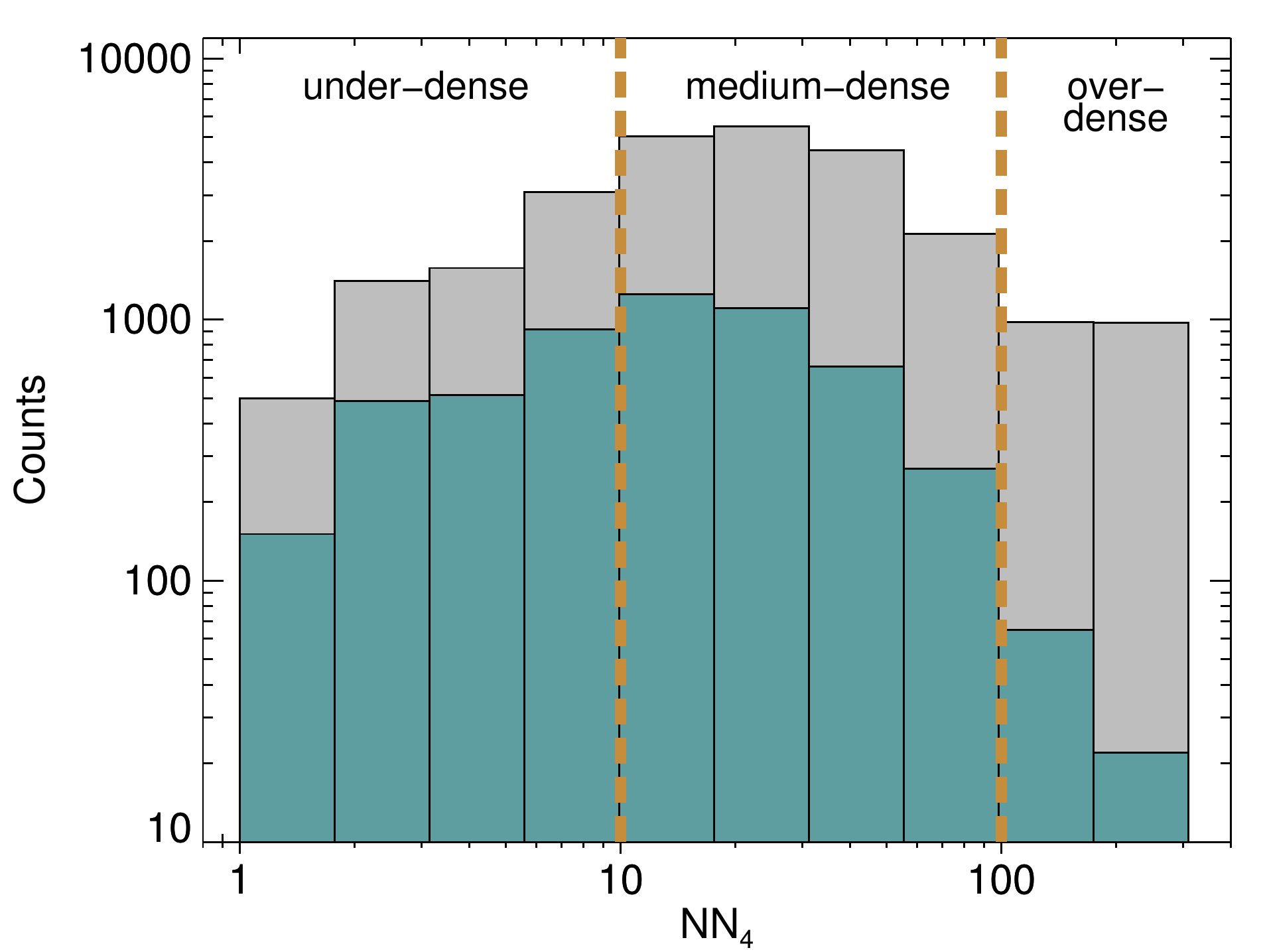}
\caption{Distribution of neighbour galaxies, NN$_4$, for all the SDSS galaxies with stellar mass of log ${\rm M}_*$/${\rm M}_{\odot}$ > 9.6 (grey histogram) and the primary galaxies with stellar mass of 10.0 < log ${\rm M}_*$/${\rm M}_{\odot}$ < 10.5 (blue histogram). According to the NN$_4$ value, we divide the total distribution into three environmental regions using the cut of 10 and 100 which are indicated by two vertical dashed lines, and label them with "under-dense", "medium-dense" and "over-dense" respectively.} 
\label{fig:env}
\end{figure}

Environment can have strong effects on galaxy evolution. It has been discovered on small scales, i.e. within the virial radius, that the evolution of satellite galaxies depends strongly on their environment (e.g. \citealt{Gunn1972}, \citealt{Butcher1978}, \citealt{Dressler1980}, \citealt{Larson1980}, \citealt{Moore1996}). On larger scales, such an effect is not yet clearly established (e.g. \citealt{Blanton2009}). There are several different ways to define environment in the literature, including the halo mass, distance to the  $k$th-nearest neighbour galaxy, number density within certain radius, etc. Here we define the environment, NN$_4$, as the total number of neighbours between a projected distance of 500 kpc and 4 Mpc from each given galaxy, and for which  the velocity difference is 1000 km/s, in accordance with our isolation criterion described in section \ref{sec:observ_data}.  In Fig.~\ref{fig:env}, the blue histogram shows the distribution of NN$_4$ for primary galaxies with stellar mass  in the range 10.0 < log ${\rm M}_*$/${\rm M}_{\odot}$ < 10.5. In this mass range, K13 found a strong conformity signal up to 4 Mpc, which could hardly be explained by the one-halo term environmental effect since their viral radii are far smaller than 4 Mpc. In the following we focus on this particular mass range, unless stated otherwise. Compared with the NN$_4$ distribution of all galaxies (grey histogram), it drops significantly at the high NN$_4$ end, indicating a large fraction of satellite galaxies in this regime.

\begin{figure}
\centering
\includegraphics[width=\linewidth]{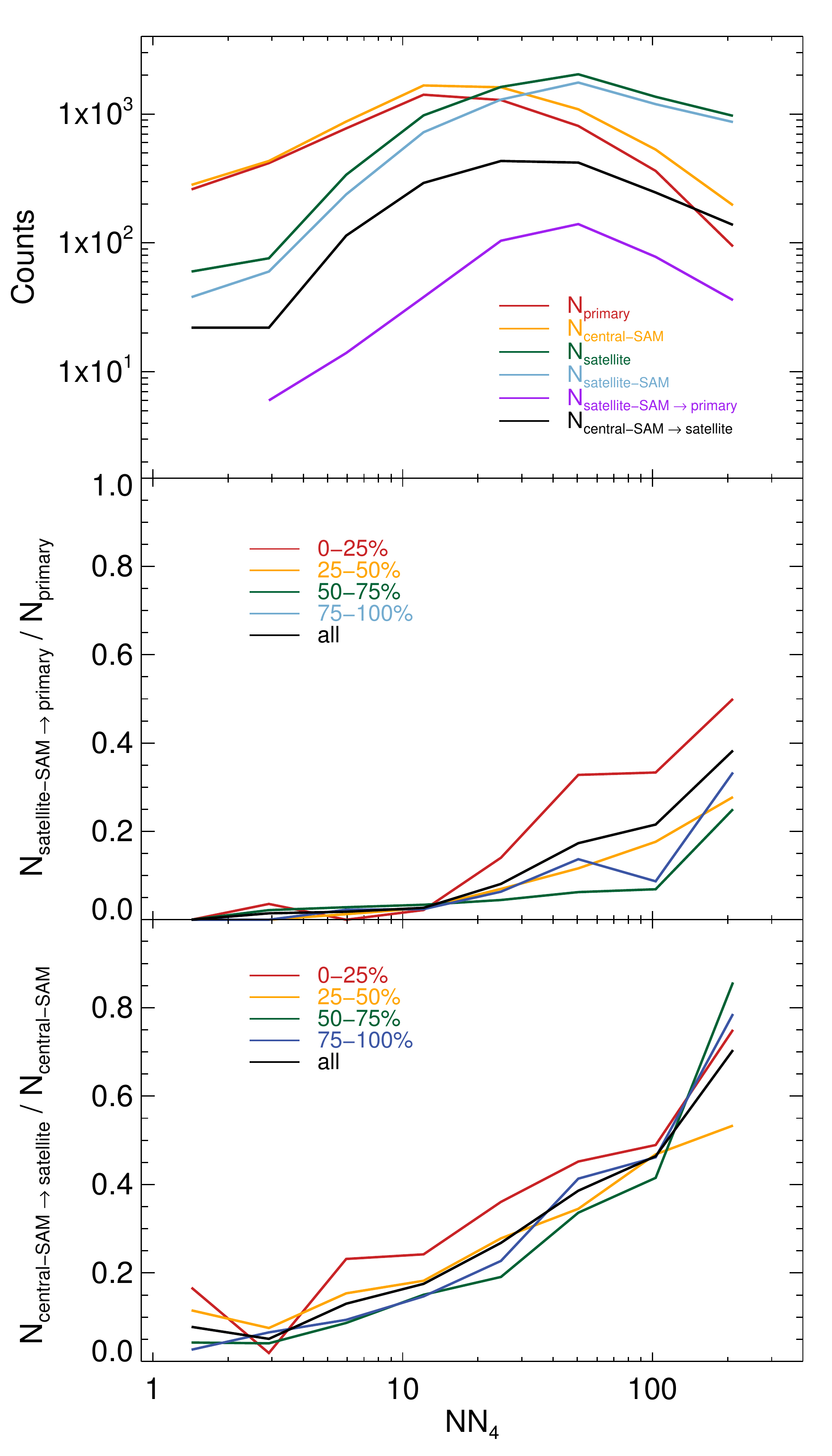}
\caption{
Top panel: Numbers of different types of galaxies as a function of NN$_4$. Red, orange, green, blue purple and black lines indicate the distribution of primaries, central-SAM galaxies, satellites, satellite-SAM galaxies, primaries which are actually satellite-SAM galaxies and satellites which are actually central-SAM galaxies, respectively. Middle panel: Fractions of satellite-SAM among primaries. Bottom panel: Fractions of central-SAM galaxies misclassified as satellite galaxies. The primaries and central-SAM galaxies are both divided into four quartiles according to their sorted sSFR values. Red, orange, green, blue and black lines indicate the 0-25\%, 25-50\%, 50-75\%, 75-100\% quartiles with increasing sSFR, respectively. Black curves show results for the full primary (middle panel) and central-SAM (bottom panel) sample. All results are for primary galaxies with stellar mass  of 10.0 < log ${\rm M}_*$/${\rm M}_{\odot}$ < 10.5 and their neighbour galaxies.
}
\label{fig:contamination}
\end{figure}

Given a certain isolation criterion, the purity and completeness of selected central galaxies could be highly correlated with their neighbour density. It could misclassify some satellite-SAM galaxies as primary galaxies, and central-SAM galaxies as satellite galaxies, especially at high NN$_4$. Using the model galaxy catalogue, we can quantify both the purity and the completeness of the primary galaxies. With the current selection criterion, we show in the upper panel of Fig.~\ref{fig:contamination} the numbers of primaries, central-SAM galaxies, satellites, satellite-SAM galaxies, primaries which are actually satellite-SAM galaxies and satellites which are actually central-SAM galaxies as functions of NN$_4$. 
It demonstrates that with the current selection criterion, the missed number of central-SAM galaxies is indeed larger than the number of false primary galaxies, i.e. satellite-SAM galaxies that are misclassified as primary galaxies. 

The middle panel shows that the contamination of primary galaxies (fraction of primary galaxies which are actually satellite-SAM, N$_{\rm satellite-SAM\rightarrow primary}$ / N$_{\rm primary}$) is an increasing function of NN$_4$. It is below 5\% at NN$_4 <$ 20, and reaches above 20\% at NN$_4$ $\sim$ 100. The incompleteness of central-SAM galaxies (fraction of central-SAM galaxies which are classified as satellite galaxies based on the isolation criterion, N$_{\rm central-SAM \rightarrow satellite}$ / N$_{\rm central-SAM}$) is also an increasing function of NN$_4$ (bottom panel). It varies from $\sim$ 5\% at NN$_4 \sim$ 3 and reaches $\sim$ 50\% at NN$_4 >$ 100. The results could thus be affected more seriously by the incompleteness than by the contamination. 

The contamination effects are even more obvious when splitting the primaries into 4 quartiles according to their sSFR values. For the most quiescent quartile in very high density regions, the contamination can be as high as 50\%,  consistent with \cite{Tinker2017,Sin2017,Zu2018}. For the most active quartile, around 20\% of primary galaxies are actually satellite-SAM galaxies. The incompleteness of central-SAM galaxies, however, does not have strong dependence on sSFR. In high density regions, more than half of central-SAM galaxies are classified as satellite galaxies both for active and for passive galaxies. Given the fact that it is difficult to simultaneously isolate true primary galaxies and guarantee the completeness of the sample of central galaxies, and the fact that purity and completeness of central galaxies are highly correlated with the neighbour abundance NN$_4$, we will not explore further selection effects but focus on the environmental effects on conformity in the following. 

\section{Results} 

In this section, we first study how the conformity signal varies in different environments. We then present the dependence of the conformity signal on stellar masses of primary galaxies. Finally we compare results to predictions from the galaxy formation models.

\subsection{Environmental effect on galaxy conformity} 
\label{sec:cnfm_env}

\begin{figure*}
\centering
\includegraphics[width=1.0\linewidth]{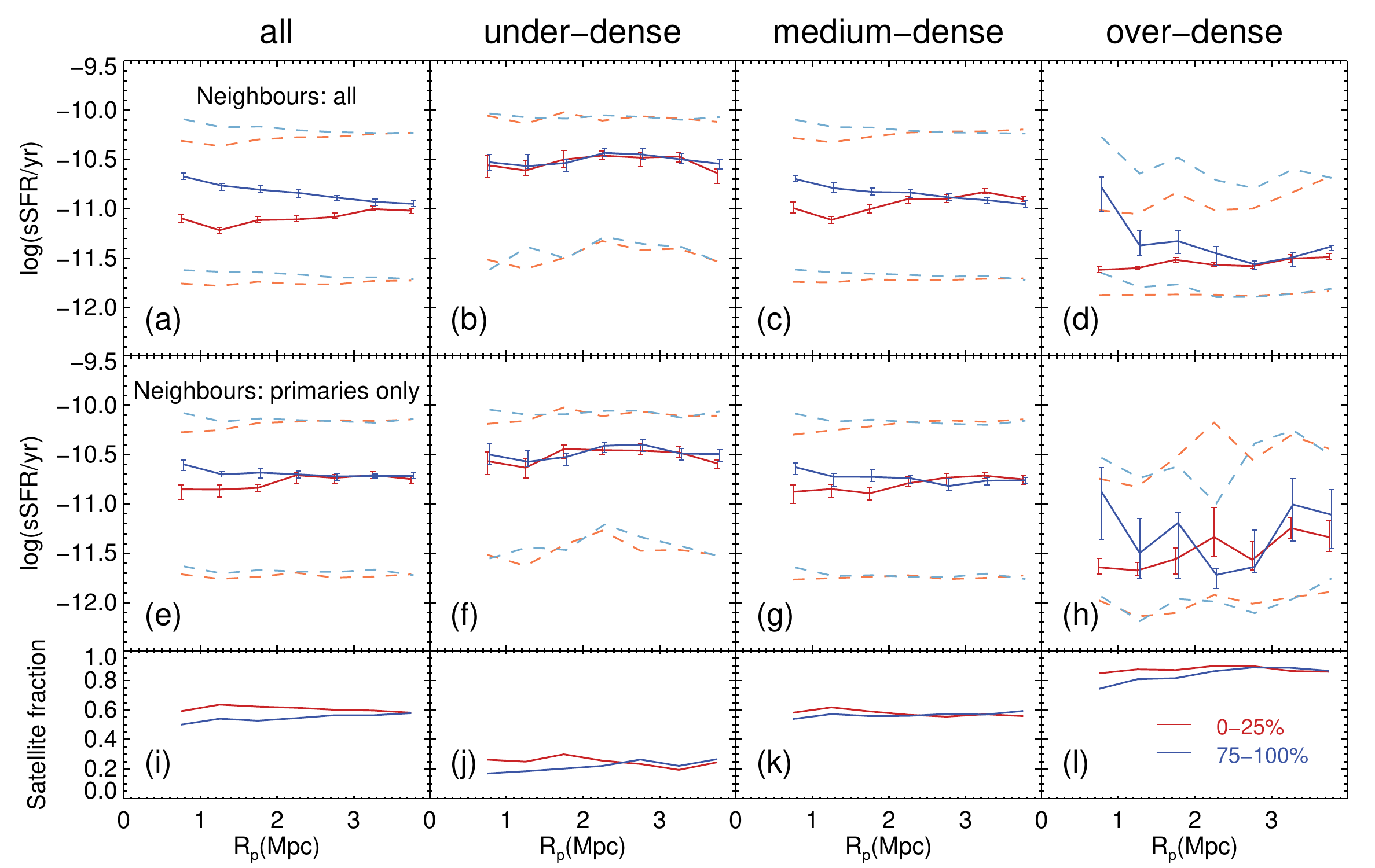}
\caption{For primary galaxies of 10.0 < log ${\rm M}_*$/${\rm M}_{\odot}$ < 10.5, the sSFR of their neighbour galaxies (top panels) or their primary neighbour galaxies (middle panels) and the satellite fraction (bottom panels) among their neighbours as a function of projected distance. The leftmost column shows the results for neighbours around all the primaries. The other three columns show results for those around primaries residing in under-dense, medium-dense and over-dense regions, respectively. The red solid curves indicate the median sSFR for neighbour galaxies around the primaries with the lowest 25\% sSFR, while the blue solid curves are for those around the primaries with the highest 25\% sSFR. Error bars show the error on the median value calculated using the bootstrap method. The lower and upper dashed curves in each panel indicate the 25th and 75th percentiles of the sSFR distributions, respectively. The blue curves are shifted horizontally by 0.03 dex to avoid  overlap of  error bars.}
\label{fig:cnfm_dens}
\end{figure*}

Panel (a) in Fig.~\ref{fig:cnfm_dens} shows the sSFRs of neighbour galaxies as a function of their projected distance from the primaries. The primary samples are split into four quartiles according to their sSFR values and we focus on the highest and the lowest ones to highlight the difference. Red curves are results for the primary galaxies that have the lowest 25\% sSFR  which we refer to as "passive primaries" hereafter. Blue curves show results for primaries with the highest 25\% sSFR, which are referred to as "active primaries". 
Solid curves give the median sSFR of neighbour galaxies, while dashed curves indicate the corresponding 25th and 75th percentiles. Consistent with the result of K13,  the most passive primaries are surrounded by passive neighbours, while active primaries are surrounded by active neighbours. The conformity signal is strong at scales between 0.7 and 3 Mpc, and becomes weaker at scales larger than $\sim$3 Mpc. Compared with the result shown in the lower left panel of Figure 2 in K13, the signal strength is slightly weaker, which can be caused by many factors, including a more strict isolation criterion, a higher stellar mass limit, different definition of SFR, etc. 

We divide the primaries into three regions according to the number of neighbour galaxies and investigate how the conformity varies in different regions. We label primary galaxies with NN$_4 < 10$ to be in "under-dense region",  the ones with  $10 <$ NN$_4 <$ 100 to be in "medium-dense region", and the ones with NN$_4 >$ 100 to be in "over-dense region". The under-dense, medium-dense and over-dense regions  cover 40\%, 58\% and 2\% of the primary galaxies, respectively. As shown in Fig.~\ref{fig:cnfm_dens}, in  under-dense regions the median sSFR of neighbours is high, regardless the activity of the primary galaxies, while for those in over-dense regions, neighbour galaxies tend to be passive, regardless of the activity of primary galaxies. In the medium-dense regions, the conformity signal is reduced and only exists within 2.5 Mpc.

\begin{figure}
\centering
\includegraphics[width=\linewidth]{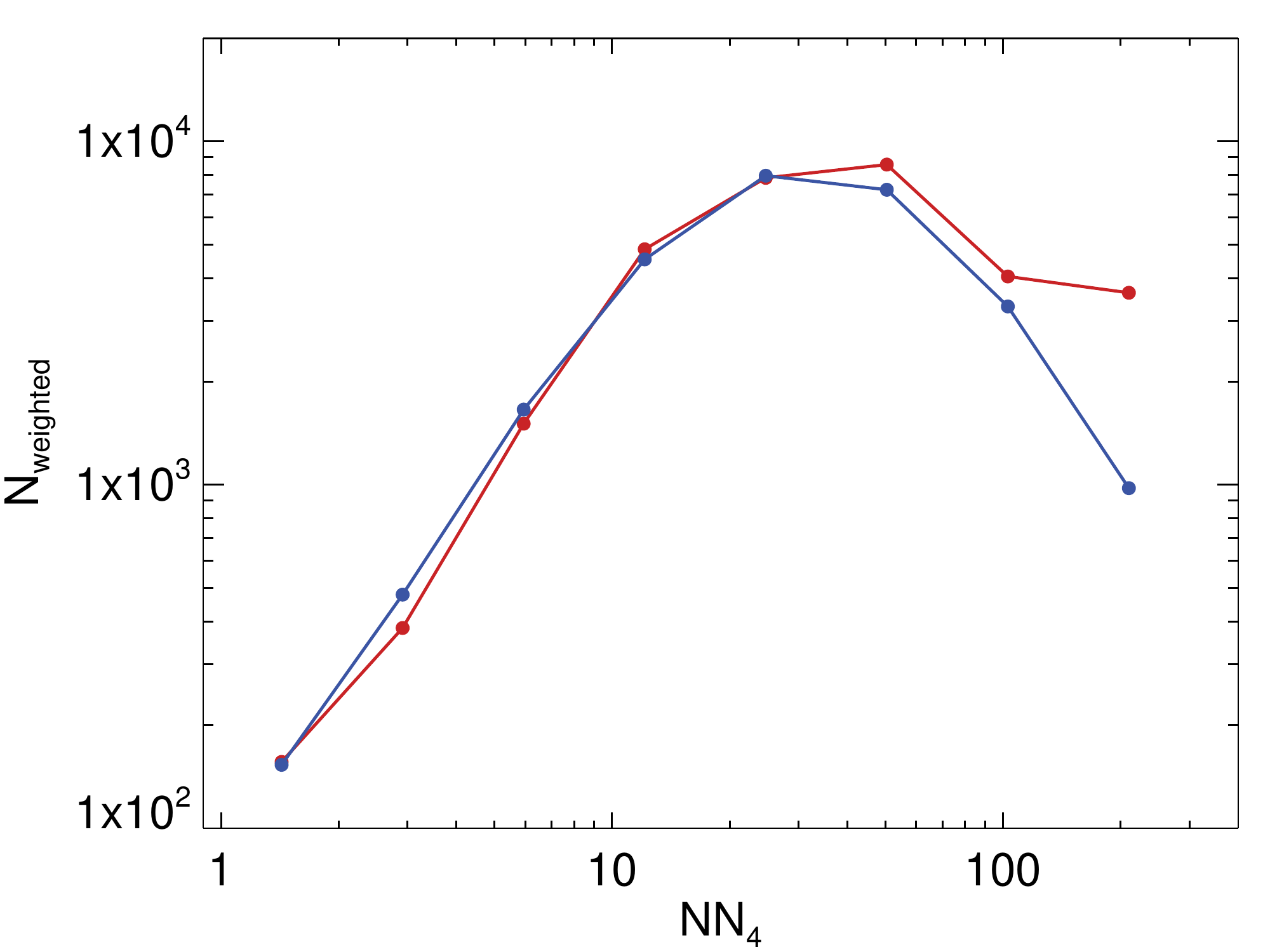}
\caption{Distribution of primary galaxies weighted by NN$_4$ as a function of NN$_4$. The red and blue curves indicate the most passive and active primaries.}
\label{fig:weightN}
\end{figure}

The difference of the sSFR of neighbour galaxies in the three density environments is much larger than the conformity signal in each individual environment, suggesting that it may be the different distribution of blue and red primary galaxies within different density environments that leads to the apparent strong signal of conformity for the full sample. Note that in this analysis, neighbour galaxies of each primary galaxy are stacked together to get a median value of sSFR of neighbours. Primaries with more neighbours thus contribute more when estimating the median sSFR of neighbours. We weight each primary galaxy by its NN$_4$ and plot the distribution of the active and passive primaries as a function of NN$_4$ in Fig.~\ref{fig:weightN}. At NN$_4 <$ 20, the distribution of the active and passive primaries are quite similar, while at NN$_4 >$ 20, there are many more passive primaries. In combination with the fact that in over-dense regions, neighbour galaxies tend to be more passive, this could lead to a rather strong signal of conformity. The effect of the stacking will be further investigated in Sec ~\ref{sec:no_stacking}.

The next question is which population of neighbour galaxies dominates the conformity signal. Central galaxies are usually less affected by environments than satellite galaxies ~\citep[e.g.][]{Peng2010}. We thus firstly explore whether there is a conformity signal carried by the primary neighbour galaxies. Panel (e) of Fig. ~\ref{fig:cnfm_dens} shows that the conformity signal is reduced significantly after removing the satellite neighbour galaxies, suggesting that it is the population of satellite galaxies that dominate the conformity signal. The primary neighbour galaxies tend to be star-forming both in the under-dense and medium-dense regions, which cover 98\% of the primary galaxies. In the over-dense region, the primary neighbours tend to be passive. Yet in this region, the possibility of misclassification of satellite-SAM as primary galaxies is high. The estimated median sSFR could have been higher after removing the misclassified population. The primary neighbour galaxies only play a minor role in the large-scale conformity.

Compared to central galaxies, satellite galaxies are in general more passive, due to starvation,  stripping, depletion of available gas supply etc (e.g. \citealt{Gunn1976}, \citealt{Abadi1999}, \citealt{Quilis2000}). A higher satellite fraction could result in an overall more passive neighbour population. Panel (i) of Fig. ~\ref{fig:cnfm_dens} shows that the satellite fraction around passive primaries is higher than that around active galaxies by $~\sim$ 10\%. This fraction is a strong function of environments. In  over-dense regions, the satellite fraction can be as high as 90\%. Given the fact that in this region, the misclassification of satellites as primary galaxies can reach 60\%, the "true" satellite fraction could have been even higher. On the contrary, in under-dense regions, the satellite fraction is only about 20\% and the neighbour galaxies are dominated by primary neighbour galaxies. In the medium-dense regions, this fraction is around 60\%.  Similar to the conformity signal, the difference of satellite fraction values in different environments are much larger than the difference of satellite fraction around passive and active galaxies in any given environment. It is the density dependent satellite fraction that dominates the median sSFR of neighbour galaxies in different environments.

\begin{figure*}
\centering
 \includegraphics[width=0.7\linewidth]{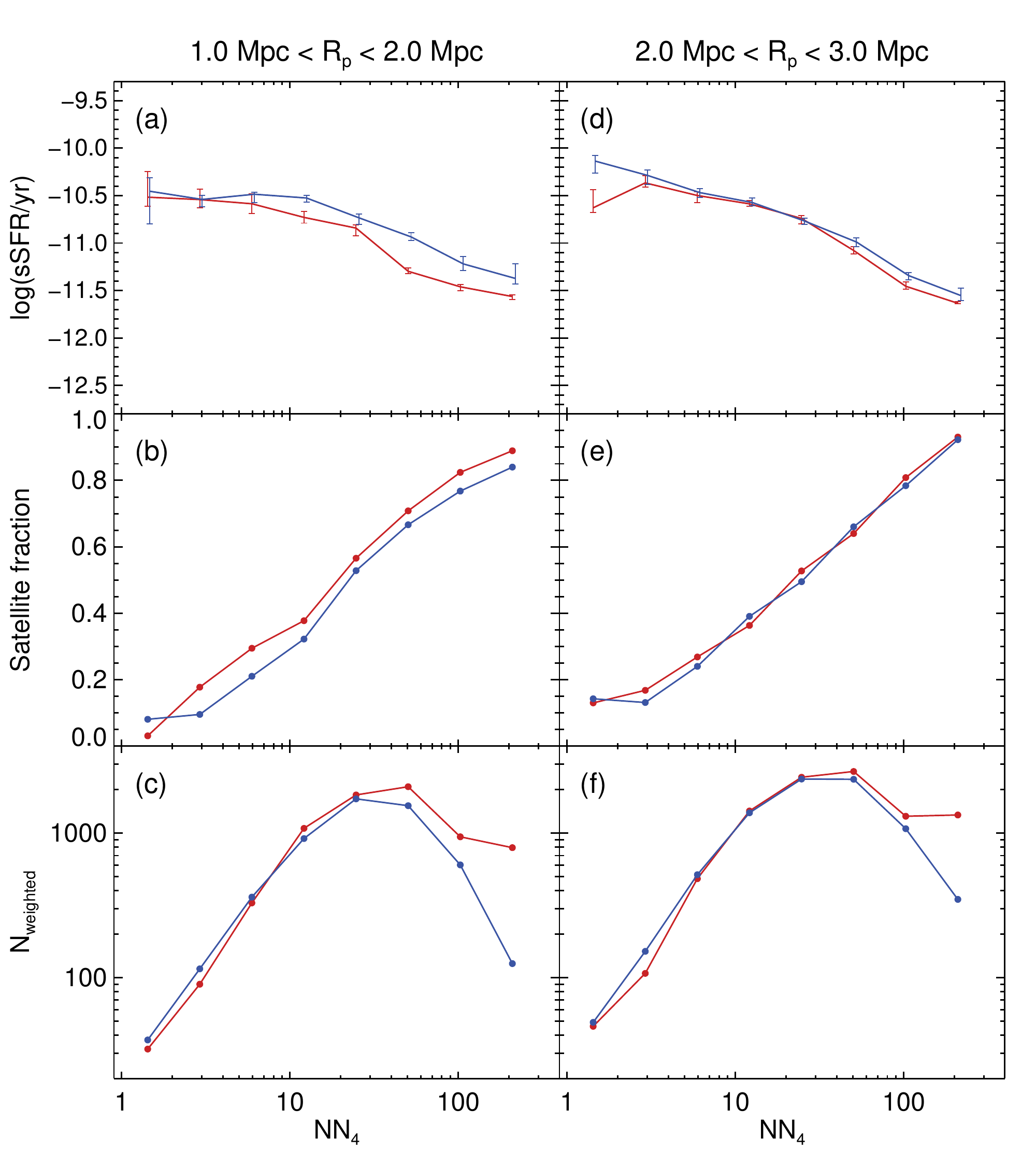}
\caption{
Top: The median sSFR of neighbours as a function of NN$_4$ around active and passive primary galaxies. 
Middle: Satellite fractions of neighbour galaxies. 
Bottom: The number of primary galaxies weighted by their neighbour galaxies within a certain distance range. 
The left panels are results for the neighbours with a projected distance of 1-2 Mpc from their primary galaxies, while right panels are for those with a projected distance of 2-3 Mpc. Blue and red curves are for neighbour galaxies around active and passive galaxies, respectively.
}
\label{fig:abund_ssfr}
\end{figure*}

To see the environmental effect in a  more continuous way, in Fig.~\ref{fig:cnfm_dens} we investigate the median sSFR distribution as a function of NN$_4$ for neighbour galaxies within the distance of 1-2 Mpc from the primaries, where the conformity signal is the strongest. As in Fig.~\ref{fig:cnfm_dens}, we find in the panel (a) of Fig.~\ref{fig:abund_ssfr} that the specific star formation rate  decreases with NN$_4$ rapidly. The difference of sSFR around the active and passive primary galaxies ($\leq$ 0.3 dex) is much smaller than the difference induced by the difference in NN$_4$ ($\sim$ 1 dex). Neighbours in low NN$_4$ regions are mostly active galaxies, while at NN$_4$ $>$ 40, regardless of the sSFR of the primary galaxies, neighbour galaxies are dominated by passive galaxies which are mostly satellite galaxies. As we can see in  panel (b), satellite fractions increase rapidly with increasing NN$_4$ value. Besides, there are more passive primary galaxies at high NN$_4$ when weighted by the number of neighbours within the distance range of 1-2 Mpc, as shown in panel (c).

\begin{figure}
\centering
\includegraphics[width=\linewidth]{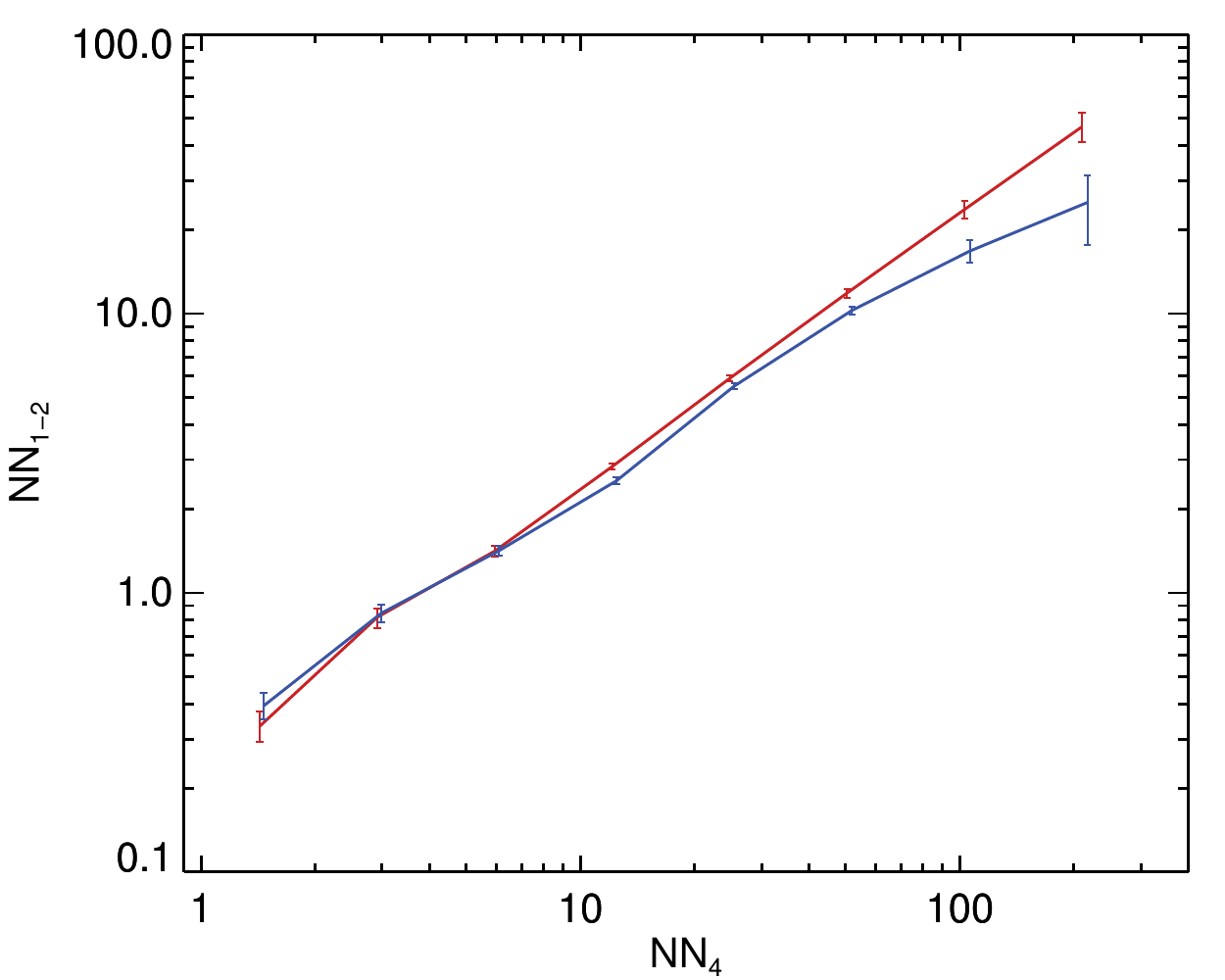}
\caption{
The relation between NN$_{1-2}$ and NN$_4$. 
The red curve shows the mean values for neighbour galaxies around passive primary galaxies, while the blue curve shows results for neighbour galaxies around active primary galaxies. Error bars on the mean values are calculated using the bootstrap method.
}
\label{fig:numdensity_abund}
\end{figure}

This residual conformity signal increases slightly with NN$_4$. At NN$_4$ smaller than $\sim$10, there is no conformity signal at all. At larger  NN$_4$, the conformity strength is around $\sim$ 0.3 dex. The residual signal of conformity can be explained by the difference in satellite fraction and the difference of local density.  As shown in panel (b) of Fig.~\ref{fig:abund_ssfr}, the satellite fraction is slightly higher for red primaries compared to the blue primaries. The local density also plays a role in the residual conformity signal. As shown in Fig. ~\ref{fig:numdensity_abund}, for a given NN$_4$, the local density (the number of neighbour galaxies within 1-2 Mpc projected distance from the primary galaxies, NN$_{1-2}$)  for red primaries is higher than for blue primaries and the difference increases with NN$_4$. At higher local density, galaxies tend to be redder, especially for satellite galaxies. In combination, the higher satellite fraction and the higher local density around passive primary galaxies lead to a slightly more passive neighbour galaxy population compared to those around the active primary galaxies. 

We apply the same analysis to neighbour galaxies 2-3 Mpc away from the primary galaxies and find that the conclusion is similar, except that the difference in satellite fraction is even smaller and the conformity signal is smaller as well.

In summary, passive primary galaxies tend to reside in over-dense regions which are dominated by passive neighbours made mostly of satellite galaxies, while active primary galaxies tend to reside in relatively under-dense regions which are dominated by active neighbours made mostly of primary neighbour galaxies. It is the combination of the different spatial distribution between passive and active primary galaxies and their environmental dependent neighbour satellite fraction that leads to the apparently strong conformity signal.

\subsection{Number weighting effect} 
\label{sec:no_stacking}

\begin{figure*}
\centering
\includegraphics[width=1\linewidth]{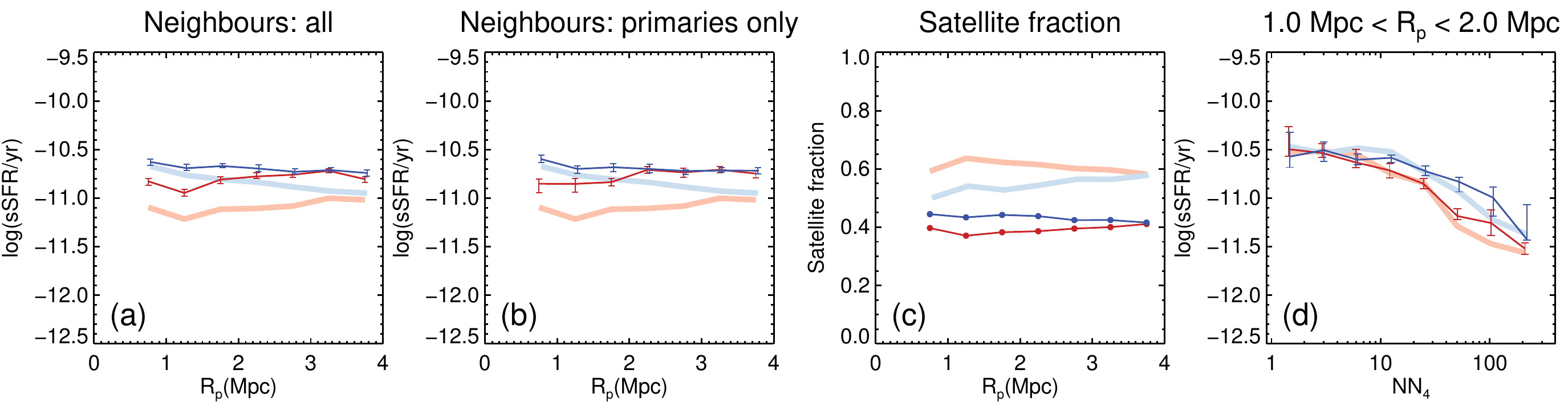}
\caption{ 
Results with the EW method, where each primary system contributes equally to the final result. Panel (a) shows the conformity signal including all neighbour galaxies; panel (b) shows the "primary neighbour only" conformity, where only primary type galaxies are included as neighbours; panel (c): satellite fraction with the EW method; panel (d): median sSFR of neighbours with projected distance of 1-2 Mpc from their primaries as a function of NN$_4$. 
In all panels, the thin curves with error bars are results for the EW method, while thick light curves show results of the NEW method as reference. Red curves are for neighbours around passive primaries, while the blue solid ones are for neighbours around active primaries. 
}
\label{fig:remove_stack}
\end{figure*}

The method used in the previous section for calculating the median neighbour sSFR could amplify the conformity signal, since primaries in dense regions are given higher weights than those in low density regions. Here we investigate how strong this effect is.

We adopt an equal-weighted (EW) approach by calculating the final median neighbour sSFR based on the median neighbour SFR estimated in each primary-neighbour system, so that each system contributes only once. The left panel of  Fig.~\ref{fig:remove_stack} shows that the conformity signal (thin curves with error bars)  is significantly reduced compared to results using the previous (non-equal-weighted, NEW) method (thick light curves). This is consistent with what found by \cite{Sin2017}. In this case, the conformity signal carried by primary neighbour galaxies and that carried by all neighbour galaxies are very similar. The satellite fraction of neighbour galaxies is estimated in a similar manner and is now lower around passive than around the active primaries. The conformity signals is still present at scales $<2$ Mpc. The last panel shows that the dependence of the median neighbour sSFR on NN$_4$ and the dependence on the activity of the primary galaxies with the EW method are both similar with those measured with the NEW.

\subsection{Dependence of conformity on primary stellar mass} 
\label{sec:cnfm_mass}

\begin{figure*}
\centering
\includegraphics[width=0.7\linewidth]{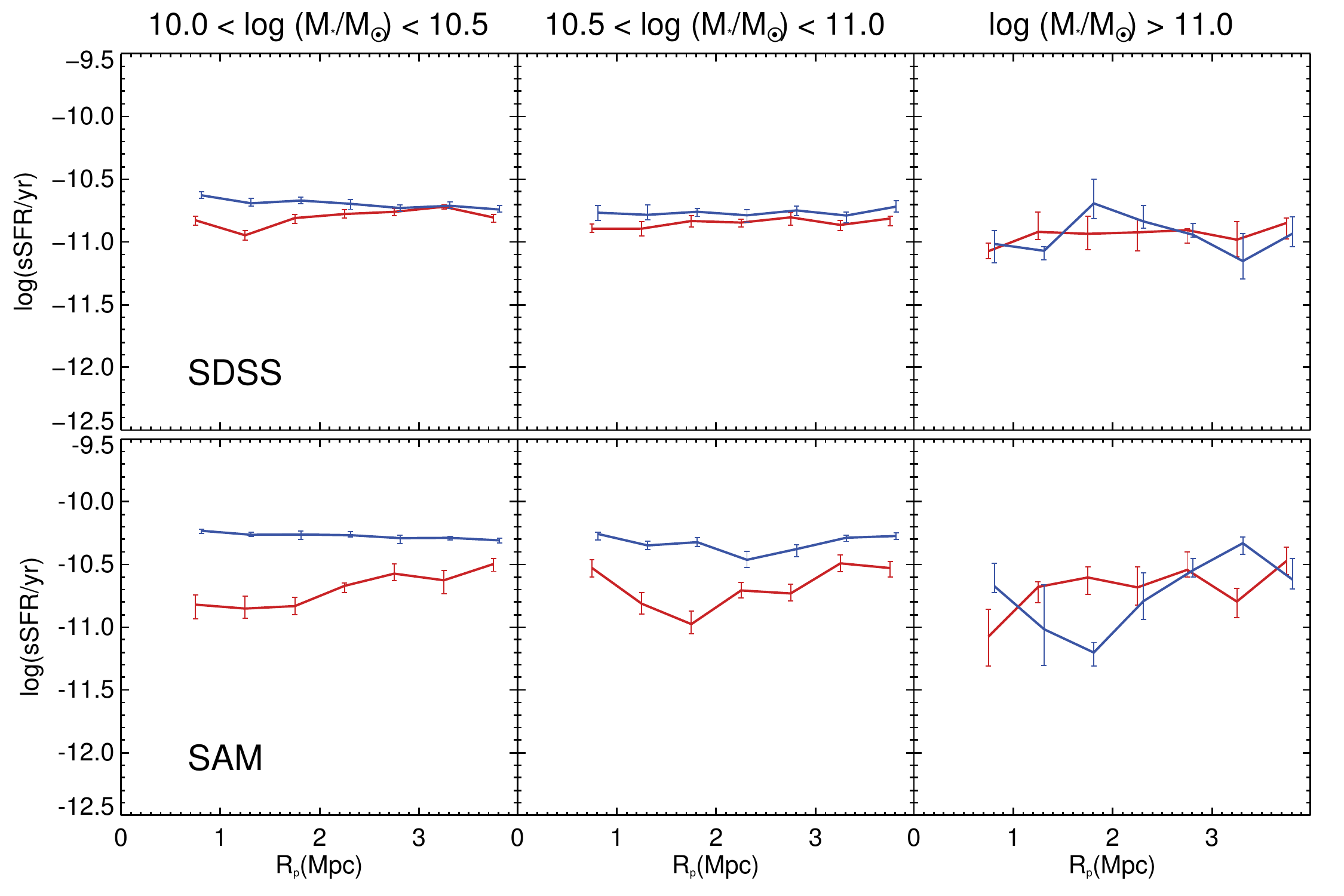}
\caption{
The EW conformity for primary galaxies in three stellar mass bins:  10.0 < log ${\rm M}_*$/${\rm M}_{\odot}$ < 10.5,  10.5 < log ${\rm M}_*$/${\rm M}_{\odot}$ < 11.0 and  log ${\rm M}_*$/${\rm M}_{\odot}$ > 11.0. The upper panels show results for the SDSS data, and the lower panels show the results for the SAM galaxies.
}
\label{fig:cnfm_mass}
\end{figure*}

In this subsection, we check how the equal-weighted conformity signals vary with the stellar mass of primary galaxies. Apart from the mass range of  10.0 < log ${\rm M}_*$/${\rm M}_{\odot}$ < 10.5 that we focused on previously, we also look at results for primary galaxies with 10.5 < log ${\rm M}_*$/${\rm M}_{\odot}$ < 11.0 and for stellar mass larger than $10^{11}M_{\odot}$. Results are shown in the upper panels of Fig.~\ref{fig:cnfm_mass}. We find the conformity signal becomes weaker as stellar mass increases and vanishes in the most massive range. This trend is consistent with the result from K13.

\subsection{Conformity in semi-analytic models} 
\label{sec:cnfm_ms2}

\begin{figure*}
\centering
\includegraphics[width=1\linewidth]{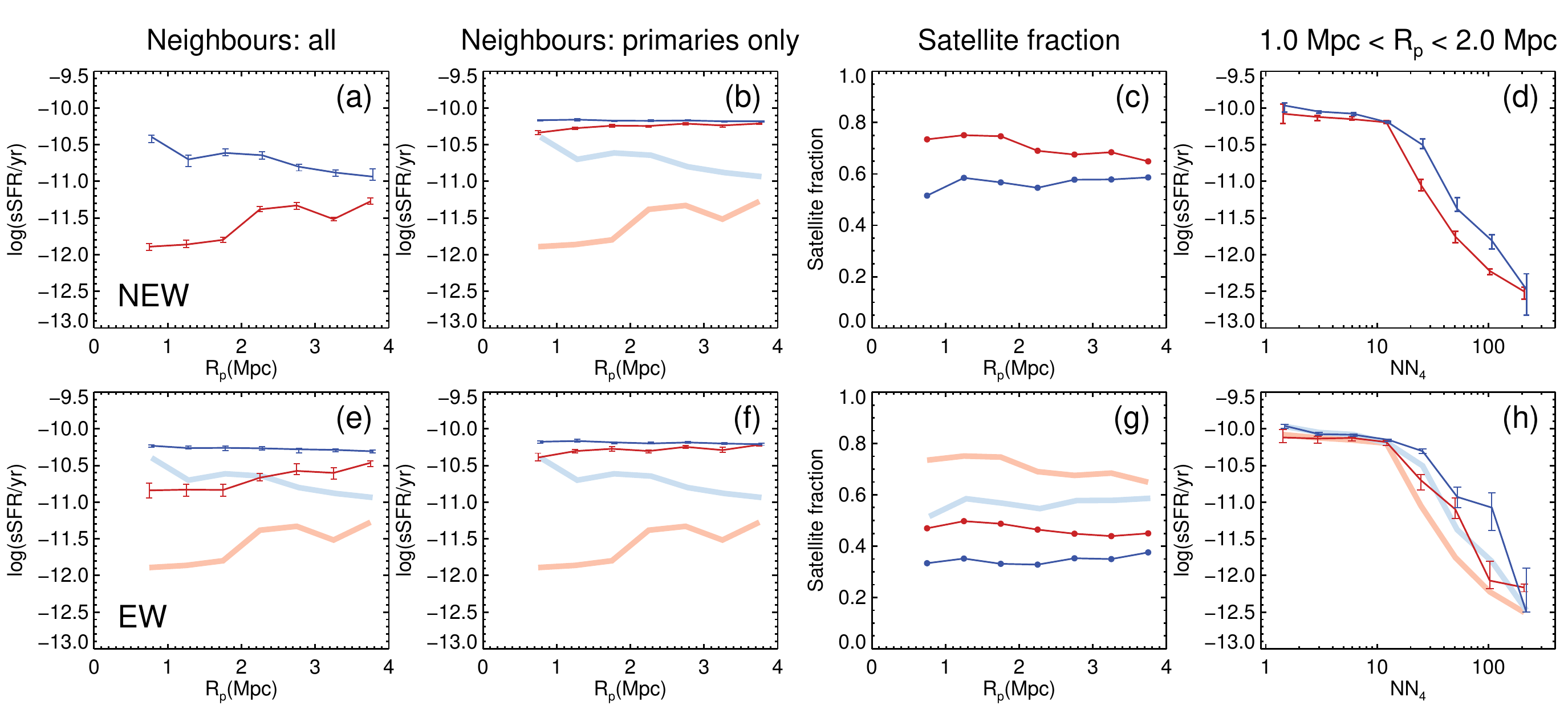}
\caption{Galactic conformity for semi-analytical galaxies. The upper panels show results measured with the NEW method, while the bottom panels show results measured with the EW method. Panel (a) and (e): conformity as a function of projected distance to their primary galaxies for all neighbour galaxies; panel (b) and (f): conformity as a function of projected distance to their primary galaxies for primary neighbour galaxies; panel (c) and (g): satellite fractions as a function of projected distance to their primary galaxies; panel (d) and (h): conformity signal as a function of NN$_4$. Blue and red curves are for neighbour galaxies around active and passive galaxies, respectively.
The thin solid lines in panel (a) are redrawn as the thick light lines in panel (b), (e) and (f) as reference. Similarly, the thin solid lines in panel (c) and (d) are redrawn as the thick light lines in panel (g) and (h), respectively.
}
\label{fig:cnfm_ms2}
\end{figure*}

In this subsection, we re-examine the conformity in the semi-analytic models with the same processing methods as for the observational data. 

Fig.~\ref{fig:cnfm_ms2} shows the results both with the NEW method (upper panels) and with the EW method (lower panels). We find in panel (a) that the model predicted conformity signal is strong, much stronger than that in observations. This conformity can be reduced significantly when  taking into account only primary neighbour galaxies (panel (b)). These results are similar to those found by \cite{Sin2017} though the sample and selection are slightly different. Note that the predicted conformity of primary neighbour galaxies is close to observations, indicating that it is the over abundance of passive satellite galaxies that leads to the strong conformity signal found in the model. Such passive satellite galaxies usually reside in high density environments and the high abundance could have a large weight when using the NEW method. With the EW method, contributions from such passive neighbour satellite galaxies can be reduced significantly. As a result, the conformity signal is reduced by about a factor of 2 (panel (e)). 

Another factor that could amplify the conformity signal is the treatment of extremely passively galaxies in the semi-analytic models. In the simulated galaxy catalogue, the SFR of some galaxies are set to be 0, corresponding to quiescent galaxies with very low SFR in observations. To get a more fair comparison, we set those galaxies to have log sSFR[yr$^{-1}$]=-12.5 in our analysis. This treatment of sSFR could also boost the conformity signal in terms of the sSFR value. 

We also study the conformity as a function of NN$_4$ for neighbour galaxies within the projected distance between 1 and 2 Mpc. Panel (d) clearly shows that the star formation of neighbour galaxies with high NN$_4$ is strongly suppressed, while for those with low NN$_4$, most neighbour galaxies are star forming galaxies. As we found in observations, the conformity signal disappears at NN$_4$ $<10$, but shows up at higher NN$_4$.  The model predictions are consistent with what we find using the SDSS data, but the environmental dependence is much stronger, which leads to a rather strong conformity signal when combining samples from different environments together.

The stellar mass dependence of the conformity is shown in the lower panels in Fig. ~\ref{fig:cnfm_mass}. We find that the conformity signal becomes weaker as the primaries become more massive. For galaxies less massive than $10^{11}$M$_{\odot}$ the conformity persists, while for even more massive galaxies, it disappears. This is consistent with the results from SDSS.

\section{Conclusions} 
Galactic conformity is found to persist on scales up to 4 Mpc, even for galaxies as massive as the Milky Way, suggesting a co-evolution of well separated distinct systems. In this paper, we study the role that environmental effects play in the apparent strong large-scale galactic conformity. 

We find that large-scale conformity is closely related to environment.  In under-dense environments, neighbour galaxies are mostly star-forming, while in over-dense region, neighbour galaxies are mostly passive, regardless of the activity of the primary galaxies. The difference between the sSFR of neighbours around the active and passive primary galaxies at fixed NN$_4$ environment is much smaller than the difference caused by the  NN$_4$ environments. We also find that  the dependence of satellite fraction on environments is strong. In low density regions, the satellite fraction of neighbour galaxies is only around 20\%, while in high density regions, satellite galaxies dominate the neighbour populations. In combination with the fact that satellite galaxies tend to be redder, this leads to  redder neighbour galaxies around primary galaxies in higher density regions. If restricted to primary neighbour galaxies, the conformity signal reduces dramatically. The apparent strong signal of conformity is due to the combination of over-abundance of passive primary galaxies in high NN$_4$ regions and the high fraction of satellite galaxies in these high NN$_4$ regions.

When split into different NN$_4$ environments, we find that the residual conformity is an increasing function of NN$_4$. In regions where NN$_4$ $<$ 10, there is no conformity signal at all, whereas at higher NN$_4$, the difference in sSFR around passive and active primaries can reach 0.2 dex at a scale of 1-2 Mpc where the conformity signal is the strongest. This is because even with the same NN$_4$, the local density estimated by the number count of neighbours within 1-2 Mpc from the passive primaries are still higher, in which both primary galaxies and satellite galaxies are redder. In addition, the satellite fraction is slightly higher around passive primary galaxies than  around active primary galaxies. A higher fraction of satellites could also lead to a lower median sSFR.

As pointed by \cite{Sin2017}, the conformity signal could be amplified if systems with more neighbours contribute more to the final median sSFR.  When we remove this effect by requiring each primary to contribute equally to the median neighbour sSFR, we find that the conformity signal is reduced by a factor of two, similar to what was found by \cite{Sin2017}. The residual conformity persist at $<$ 2.5 Mpc, and increases with NN$_4$. Such effect is highly degenerate with the dependence on NN$_4$ that we discovered. 

We find that the conformity signal is a decreasing function of the stellar mass of the primary galaxy. For those of stellar mass $>10^{11}$M$_{\odot}$, the conformity effect disappears, as in K13.

When applying the same analysis to semi-analytical model galaxies, we find that the model predictions are in reasonable agreement with those from the observational data, indicating that there is no need to include new physics to reproduce such a large-scale conformity and its dependence on the stellar mass of the primary galaxy.

\section*{Acknowledgements}

We acknowledge support from the National Key Program for Science and Technology Research and Development (2017YFB0203300).
SS is grateful for the helpful discussions and suggestions with Prof. Cheng Li and his hospitality during a visit to the Shanghai Astronomical Observatory.
QG and LG acknowledges support from NSFC grant (No. 11425312), and two Royal Society Newton Advanced Fellowships, as well as the hospitality of the Institute for Computational Cosmology at Durham University. 
QG is also supported by two NSFC grants (Nos. 11573033, 11622325), and the "Recruitment Program of Global Youth Experts" of China, the NAOC grant (Y434011V01).
LW acknowledges support from the NSFC grant programme (No. 11573031).
JW acknowledges the 973 program grant 2015CB857005 and NSFC grants (Nos. 11373029, 11390372).
JP is supported by the Ministry of Science \& Technology through 973 grant of No. 2015CB857001 and the NSFC through grants of No. 11573030.
CGL acknowledges support from the UK Science and Technology Facilities Council [ST/L00075X/1].

\bibliographystyle{mnras}
\bibliography{ref-cnfm}

\begin{thebibliography}{}
\makeatletter
\relax
\def\mn@urlcharsother{\let\do\@makeother \do\$\do\&\do\#\do\^\do\_\do\%\do\~}
\def\mn@doi{\begingroup\mn@urlcharsother \@ifnextchar [ {\mn@doi@}
  {\mn@doi@[]}}
\def\mn@doi@[#1]#2{\def\@tempa{#1}\ifx\@tempa\@empty \href
  {http://dx.doi.org/#2} {doi:#2}\else \href {http://dx.doi.org/#2} {#1}\fi
  \endgroup}
\def\mn@eprint#1#2{\mn@eprint@#1:#2::\@nil}
\def\mn@eprint@arXiv#1{\href {http://arxiv.org/abs/#1} {{\tt arXiv:#1}}}
\def\mn@eprint@dblp#1{\href {http://dblp.uni-trier.de/rec/bibtex/#1.xml}
  {dblp:#1}}
\def\mn@eprint@#1:#2:#3:#4\@nil{\def\@tempa {#1}\def\@tempb {#2}\def\@tempc
  {#3}\ifx \@tempc \@empty \let \@tempc \@tempb \let \@tempb \@tempa \fi \ifx
  \@tempb \@empty \def\@tempb {arXiv}\fi \@ifundefined
  {mn@eprint@\@tempb}{\@tempb:\@tempc}{\expandafter \expandafter \csname
  mn@eprint@\@tempb\endcsname \expandafter{\@tempc}}}

\bibitem[\protect\citeauthoryear{{Abadi}, {Moore}  \& {Bower}}{{Abadi}
  et~al.}{1999}]{Abadi1999}
{Abadi} M.~G.,  {Moore} B.,   {Bower} R.~G.,  1999, \mn@doi [\mnras]
  {10.1046/j.1365-8711.1999.02715.x}, \href
  {http://adsabs.harvard.edu/abs/1999MNRAS.308..947A} {308, 947}

\bibitem[\protect\citeauthoryear{{Abazajian} et~al.,}{{Abazajian}
  et~al.}{2009}]{Abazajian2009}
{Abazajian} K.~N.,  et~al., 2009, \mn@doi [\apjs]
  {10.1088/0067-0049/182/2/543}, \href
  {http://adsabs.harvard.edu/abs/2009ApJS..182..543A} {182, 543}

\bibitem[\protect\citeauthoryear{{Blanton} \& {Moustakas}}{{Blanton} \&
  {Moustakas}}{2009}]{Blanton2009}
{Blanton} M.~R.,  {Moustakas} J.,  2009, \mn@doi [\araa]
  {10.1146/annurev-astro-082708-101734}, \href
  {http://adsabs.harvard.edu/abs/2009ARA%26A..47..159B} {47, 159}

\bibitem[\protect\citeauthoryear{{Blanton} \& {Roweis}}{{Blanton} \&
  {Roweis}}{2007}]{Blanton2007}
{Blanton} M.~R.,  {Roweis} S.,  2007, \mn@doi [\aj] {10.1086/510127}, \href
  {http://adsabs.harvard.edu/abs/2007AJ....133..734B} {133, 734}

\bibitem[\protect\citeauthoryear{{Blanton} et~al.,}{{Blanton}
  et~al.}{2005}]{Blanton2005}
{Blanton} M.~R.,  et~al., 2005, \mn@doi [\aj] {10.1086/429803}, \href
  {http://adsabs.harvard.edu/abs/2005AJ....129.2562B} {129, 2562}

\bibitem[\protect\citeauthoryear{{Boylan-Kolchin}, {Springel}, {White},
  {Jenkins}  \& {Lemson}}{{Boylan-Kolchin} et~al.}{2009}]{Boylan-Kolchin2009}
{Boylan-Kolchin} M.,  {Springel} V.,  {White} S.~D.~M.,  {Jenkins} A.,
  {Lemson} G.,  2009, \mn@doi [\mnras] {10.1111/j.1365-2966.2009.15191.x},
  \href {http://adsabs.harvard.edu/abs/2009MNRAS.398.1150B} {398, 1150}

\bibitem[\protect\citeauthoryear{{Brinchmann}, {Charlot}, {White}, {Tremonti},
  {Kauffmann}, {Heckman}  \& {Brinkmann}}{{Brinchmann}
  et~al.}{2004}]{Brinchmann2004}
{Brinchmann} J.,  {Charlot} S.,  {White} S.~D.~M.,  {Tremonti} C.,  {Kauffmann}
  G.,  {Heckman} T.,   {Brinkmann} J.,  2004, \mn@doi [\mnras]
  {10.1111/j.1365-2966.2004.07881.x}, \href
  {http://adsabs.harvard.edu/abs/2004MNRAS.351.1151B} {351, 1151}

\bibitem[\protect\citeauthoryear{{Butcher} \& {Oemler}}{{Butcher} \&
  {Oemler}}{1978}]{Butcher1978}
{Butcher} H.,  {Oemler} Jr. A.,  1978, \mn@doi [\apj] {10.1086/156640}, \href
  {http://adsabs.harvard.edu/abs/1978ApJ...226..559B} {226, 559}

\bibitem[\protect\citeauthoryear{{Davis}, {Efstathiou}, {Frenk}  \&
  {White}}{{Davis} et~al.}{1985}]{Davis1985}
{Davis} M.,  {Efstathiou} G.,  {Frenk} C.~S.,   {White} S.~D.~M.,  1985,
  \mn@doi [\apj] {10.1086/163168}, \href
  {http://adsabs.harvard.edu/abs/1985ApJ...292..371D} {292, 371}

\bibitem[\protect\citeauthoryear{{Dressler}}{{Dressler}}{1980}]{Dressler1980}
{Dressler} A.,  1980, \mn@doi [\apj] {10.1086/157753}, \href
  {http://adsabs.harvard.edu/abs/1980ApJ...236..351D} {236, 351}

\bibitem[\protect\citeauthoryear{{Gao}, {Springel}  \& {White}}{{Gao}
  et~al.}{2005}]{Gao2005}
{Gao} L.,  {Springel} V.,   {White} S.~D.~M.,  2005, \mn@doi [\mnras]
  {10.1111/j.1745-3933.2005.00084.x}, \href
  {http://adsabs.harvard.edu/abs/2005MNRAS.363L..66G} {363, L66}

\bibitem[\protect\citeauthoryear{{Gunn} \& {Gott}}{{Gunn} \&
  {Gott}}{1972}]{Gunn1972}
{Gunn} J.~E.,  {Gott} III J.~R.,  1972, \mn@doi [\apj] {10.1086/151605}, \href
  {http://adsabs.harvard.edu/abs/1972ApJ...176....1G} {176, 1}

\bibitem[\protect\citeauthoryear{{Gunn} \& {Tinsley}}{{Gunn} \&
  {Tinsley}}{1976}]{Gunn1976}
{Gunn} J.~E.,  {Tinsley} B.~M.,  1976, \mn@doi [\apj] {10.1086/154797}, \href
  {http://adsabs.harvard.edu/abs/1976ApJ...210....1G} {210, 1}

\bibitem[\protect\citeauthoryear{{Guo} \& {White}}{{Guo} \&
  {White}}{2008}]{Guo2008}
{Guo} Q.,  {White} S.~D.~M.,  2008, \mn@doi [\mnras]
  {10.1111/j.1365-2966.2007.12619.x}, \href
  {http://adsabs.harvard.edu/abs/2008MNRAS.384....2G} {384, 2}

\bibitem[\protect\citeauthoryear{{Guo}, {Cole}, {Eke}  \& {Frenk}}{{Guo}
  et~al.}{2011}]{Guo2011}
{Guo} Q.,  {Cole} S.,  {Eke} V.,   {Frenk} C.,  2011, \mn@doi [\mnras]
  {10.1111/j.1365-2966.2011.19270.x}, \href
  {http://adsabs.harvard.edu/abs/2011MNRAS.417..370G} {417, 370}

\bibitem[\protect\citeauthoryear{{Henriques}, {White}, {Thomas}, {Angulo},
  {Guo}, {Lemson}  \& {Wang}}{{Henriques} et~al.}{2017}]{Henriques2017}
{Henriques} B.~M.~B.,  {White} S.~D.~M.,  {Thomas} P.~A.,  {Angulo} R.~E.,
  {Guo} Q.,  {Lemson} G.,   {Wang} W.,  2017, \mn@doi [\mnras]
  {10.1093/mnras/stx1010}, \href
  {http://adsabs.harvard.edu/abs/2017MNRAS.469.2626H} {469, 2626}

\bibitem[\protect\citeauthoryear{{Kauffmann}}{{Kauffmann}}{2015}]{Kauffmann2015}
{Kauffmann} G.,  2015, \mn@doi [\mnras] {10.1093/mnras/stv2113}, \href
  {http://adsabs.harvard.edu/abs/2015MNRAS.454.1840K} {454, 1840}

\bibitem[\protect\citeauthoryear{{Kauffmann}, {Li}, {Zhang}  \&
  {Weinmann}}{{Kauffmann} et~al.}{2013}]{Kauffmann2013}
{Kauffmann} G.,  {Li} C.,  {Zhang} W.,   {Weinmann} S.,  2013, \mn@doi [\mnras]
  {10.1093/mnras/stt007}, \href
  {http://adsabs.harvard.edu/abs/2013MNRAS.430.1447K} {430, 1447}

\bibitem[\protect\citeauthoryear{{Knobel}, {Lilly}, {Woo}  \& {Kova{\v
  c}}}{{Knobel} et~al.}{2015}]{Knobel2015}
{Knobel} C.,  {Lilly} S.~J.,  {Woo} J.,   {Kova{\v c}} K.,  2015, \mn@doi
  [\apj] {10.1088/0004-637X/800/1/24}, \href
  {http://adsabs.harvard.edu/abs/2015ApJ...800...24K} {800, 24}

\bibitem[\protect\citeauthoryear{{Larson}, {Tinsley}  \& {Caldwell}}{{Larson}
  et~al.}{1980}]{Larson1980}
{Larson} R.~B.,  {Tinsley} B.~M.,   {Caldwell} C.~N.,  1980, \mn@doi [\apj]
  {10.1086/157917}, \href {http://adsabs.harvard.edu/abs/1980ApJ...237..692L}
  {237, 692}

\bibitem[\protect\citeauthoryear{{Moore}, {Katz}, {Lake}, {Dressler}  \&
  {Oemler}}{{Moore} et~al.}{1996}]{Moore1996}
{Moore} B.,  {Katz} N.,  {Lake} G.,  {Dressler} A.,   {Oemler} A.,  1996,
  \mn@doi [\nat] {10.1038/379613a0}, \href
  {http://adsabs.harvard.edu/abs/1996Natur.379..613M} {379, 613}

\bibitem[\protect\citeauthoryear{{Nelan}, {Smith}, {Hudson}, {Wegner}, {Lucey},
  {Moore}, {Quinney}  \& {Suntzeff}}{{Nelan} et~al.}{2005}]{Nelan2005}
{Nelan} J.~E.,  {Smith} R.~J.,  {Hudson} M.~J.,  {Wegner} G.~A.,  {Lucey}
  J.~R.,  {Moore} S.~A.~W.,  {Quinney} S.~J.,   {Suntzeff} N.~B.,  2005,
  \mn@doi [\apj] {10.1086/431962}, \href
  {http://adsabs.harvard.edu/abs/2005ApJ...632..137N} {632, 137}

\bibitem[\protect\citeauthoryear{{Peng} et~al.,}{{Peng}
  et~al.}{2010}]{Peng2010}
{Peng} Y.-j.,  et~al., 2010, \mn@doi [\apj] {10.1088/0004-637X/721/1/193},
  \href {http://adsabs.harvard.edu/abs/2010ApJ...721..193P} {721, 193}

\bibitem[\protect\citeauthoryear{{Peng}, {Lilly}, {Renzini}  \&
  {Carollo}}{{Peng} et~al.}{2012}]{Peng2012}
{Peng} Y.-j.,  {Lilly} S.~J.,  {Renzini} A.,   {Carollo} M.,  2012, \mn@doi
  [\apj] {10.1088/0004-637X/757/1/4}, \href
  {http://adsabs.harvard.edu/abs/2012ApJ...757....4P} {757, 4}

\bibitem[\protect\citeauthoryear{{Quilis}, {Moore}  \& {Bower}}{{Quilis}
  et~al.}{2000}]{Quilis2000}
{Quilis} V.,  {Moore} B.,   {Bower} R.,  2000, \mn@doi [Science]
  {10.1126/science.288.5471.1617}, \href
  {http://adsabs.harvard.edu/abs/2000Sci...288.1617Q} {288, 1617}

\bibitem[\protect\citeauthoryear{{Sin}, {Lilly}  \& {Henriques}}{{Sin}
  et~al.}{2017}]{Sin2017}
{Sin} L.~P.~T.,  {Lilly} S.~J.,   {Henriques} B.~M.~B.,  2017, \mn@doi [\mnras]
  {10.1093/mnras/stx1674}, \href
  {http://adsabs.harvard.edu/abs/2017MNRAS.471.1192S} {471, 1192}

\bibitem[\protect\citeauthoryear{{Thomas}, {Maraston}, {Bender}  \& {Mendes de
  Oliveira}}{{Thomas} et~al.}{2005}]{Thomas2005}
{Thomas} D.,  {Maraston} C.,  {Bender} R.,   {Mendes de Oliveira} C.,  2005,
  \mn@doi [\apj] {10.1086/426932}, \href
  {http://adsabs.harvard.edu/abs/2005ApJ...621..673T} {621, 673}

\bibitem[\protect\citeauthoryear{{Tinker}, {Hahn}, {Mao}, {Wetzel}  \&
  {Conroy}}{{Tinker} et~al.}{2017}]{Tinker2017}
{Tinker} J.~L.,  {Hahn} C.,  {Mao} Y.-Y.,  {Wetzel} A.~R.,   {Conroy} C.,
  2017, preprint, \href {http://adsabs.harvard.edu/abs/2017arXiv170201121T} {}
  (\mn@eprint {arXiv} {1702.01121})

\bibitem[\protect\citeauthoryear{{Vogelsberger} et~al.,}{{Vogelsberger}
  et~al.}{2014a}]{Vogelsberger2014a}
{Vogelsberger} M.,  et~al., 2014a, \mn@doi [\mnras] {10.1093/mnras/stu1536},
  \href {http://adsabs.harvard.edu/abs/2014MNRAS.444.1518V} {444, 1518}

\bibitem[\protect\citeauthoryear{{Vogelsberger} et~al.,}{{Vogelsberger}
  et~al.}{2014b}]{Vogelsberger2014}
{Vogelsberger} M.,  et~al., 2014b, \mn@doi [\nat] {10.1038/nature13316}, \href
  {http://adsabs.harvard.edu/abs/2014Natur.509..177V} {509, 177}

\bibitem[\protect\citeauthoryear{{Wang} \& {White}}{{Wang} \&
  {White}}{2012}]{Wang2012}
{Wang} W.,  {White} S.~D.~M.,  2012, \mn@doi [\mnras]
  {10.1111/j.1365-2966.2012.21256.x}, \href
  {http://adsabs.harvard.edu/abs/2012MNRAS.424.2574W} {424, 2574}

\bibitem[\protect\citeauthoryear{{Wang}, {Weinmann}, {De Lucia}  \&
  {Yang}}{{Wang} et~al.}{2013}]{Wang2013a}
{Wang} L.,  {Weinmann} S.~M.,  {De Lucia} G.,   {Yang} X.,  2013, \mn@doi
  [\mnras] {10.1093/mnras/stt743}, \href
  {http://adsabs.harvard.edu/abs/2013MNRAS.433..515W} {433, 515}

\bibitem[\protect\citeauthoryear{{Weinmann}, {van den Bosch}, {Yang}  \&
  {Mo}}{{Weinmann} et~al.}{2006}]{Weinmann2006}
{Weinmann} S.~M.,  {van den Bosch} F.~C.,  {Yang} X.,   {Mo} H.~J.,  2006,
  \mn@doi [\mnras] {10.1111/j.1365-2966.2005.09865.x}, \href
  {http://adsabs.harvard.edu/abs/2006MNRAS.366....2W} {366, 2}

\bibitem[\protect\citeauthoryear{{Weinmann}, {Kauffmann}, {van den Bosch},
  {Pasquali}, {McIntosh}, {Mo}, {Yang}  \& {Guo}}{{Weinmann}
  et~al.}{2009}]{Weinmann2009}
{Weinmann} S.~M.,  {Kauffmann} G.,  {van den Bosch} F.~C.,  {Pasquali} A.,
  {McIntosh} D.~H.,  {Mo} H.,  {Yang} X.,   {Guo} Y.,  2009, \mn@doi [\mnras]
  {10.1111/j.1365-2966.2009.14412.x}, \href
  {http://adsabs.harvard.edu/abs/2009MNRAS.394.1213W} {394, 1213}

\bibitem[\protect\citeauthoryear{{Zu} \& {Mandelbaum}}{{Zu} \&
  {Mandelbaum}}{2018}]{Zu2018}
{Zu} Y.,  {Mandelbaum} R.,  2018, \mn@doi [\mnras] {10.1093/mnras/sty279},
  \href {http://adsabs.harvard.edu/abs/2018MNRAS.tmp..276Z} {}

\makeatother
\end{thebibliography}

\label{lastpage}

\end{document}